\documentstyle[epsfig]{basi}
\def\bx{{\bf x}}
\def\bv{{\bf v}}
\def\tilt{\alpha_{ij}}
\def\tiltrt{\alpha_{r\theta}}
\def\tiltrp{\alpha_{r\phi}}
\def\tiltpt{\alpha_{\phi\theta}}
\def\feh{[{\rm Fe/H}]}
\def\fdg{\mbox{$.\!\!^\circ$}}

\begin{document}
\title[Chandrasekhar and Modern Stellar Dynamics]{Chandrasekhar and Modern Stellar Dynamics}

\author[N.W. Evans]%
       {N.W. Evans\thanks{e-mail:nwe@ast.cam.ac.uk} \\ 
        Institute of Astronomy, Madingley Rd, Cambridge, CB3 0HA, UK}
\maketitle
\label{firstpage}
\begin{abstract}
  Stellar dynamics occupied Chandrasekhar's interest for a brief
  interlude between his more prolonged studies of stellar structure
  and radiative transfer.  This paper traces the history of one of his
  ideas -- namely, that the shape of the galactic potential controls
  the orientation of the stellar velocity dispersion tensor. It has
  its roots in papers by Eddington (1915) and Chandrasekhar (1939),
  and provoked a fascinating dispute between these two great
  scientists -- less well-known than their famous controversy over the
  white dwarf stars. In modern language, Eddington claimed that the
  integral curves of the eigenvectors of the velocity dispersion
  tensor provide a one-dimensional foliation into mutually orthogonal
  surfaces. Chandrasekhar challenged this, and explicitly constructed
  a counter-example. In fact, the work of neither of these great
  scientists was without flaws, though further developments in stellar
  dynamics were to ultimately draw more on Eddington's insight than
  Chandrasekhar's.  We conclude with a description of modern attempts
  to measure the orientation of the velocity dispersion tensor for
  populations in the Milky Way Galaxy, a subject that is coming into
  its own with the dawning of the Age of Precision Astrometry.
\end{abstract}


\section{Introduction}
\label{sec:intro}

Chandrasekhar was perhaps the most influential theoretical
astrophysicist of his time.  This influence was particularly felt
through an outstanding series of research monographs that continue to
be read today. In fact, most astronomers first encounter Chandrasekhar
through the cheap Dover reprints of books like {\it Stellar
  Structure}, {\it Radiative Transfer}, {\it Hydrodynamic and
  Hydromagnetic Stability} and {\it Ellipsoidal Figures of
  Equilibrium}.  These books bristle with formulae, equations,
numerical tables, graphs and historical notes, leavened with an
immaculate prose style. They make exciting reading still today because
they contain so much classic astrophysics so lucidly explained.

In his Nobel lecture, Chandraskhar (1984) has written `` {\it There
  have been seven periods in my life. They are briefly: 1) stellar
  structure, including the theory of white dwarfs (1929-1939); 2)
  stellar dynamics, including the theory of Brownian motion
  (1938-1943); 3) the theory of radiative transfer, the theory of the
  illumination and the polarization of sunlit sky (1943-1950); 4)
  hydrodynamic and hydromagnetic stability (1952-1961); 5) the
  equilibrium and stability of ellipsoidal figures of equilibrium
  (1961-1968); 6) the general theory of relativity and relativistic
  astrophysics (1962-71) and 7) the mathematical theory of black holes
  (1974-1983).}''

So, Chandrasekhar's work on stellar dynamics occupied a brief
interlude of time.  It began in 1938 as an natural progression of his
interests in the structure and evolution of stars. This was at the
height of his famous controversy with Eddington over the fate of the
white dwarf stars. It was over by 1943, when Chandrasekhar was
commuting between the Yerkes Observatory in Chicago and the Aberdeen
Proving Ground in Maryland, working on ballistics as part of the war
effort. His research interests had moved towards radiative transfer --
the subject which Chandrasekhar himself has described as the one giving
him most satisfaction (Wali 1991).

Chandrasekhar's (1943) book {\it Principles of Stellar Dynamics} is
not as well-known or as magisterial as some of his others. The work on
dynamical friction and dynamics of star clusters has proved to be of
long-lasting value (see e.g., Heggie's article in this
issue). However, much of the book reads oddly today. There are two
long and, to modern eyes, puzzling chapters devoted to problems in
collisionless stellar dynamics, in particular, galaxy models
consistent with the ellipsoidal hypothesis. This term is not much used
nowadays, but was introduced by Eddington (1915) as a generalisation
of the triaxial Gaussian distribution of velocities used by
Schwarzschild (1908) to describe the velocities of stars in the solar
neighbourhood. This is the work we shall examine here, and it is fair
to say that this is not Chandrasekhar at his most memorable. But, its
connection with the earlier work of Eddington is fascinating,
especially considering the personal relations between these two great
scientists. And even when Chandrasekhar was not at his brilliant best,
he could still find much of interest that others had overlooked.

So, we shall trace out the twists and turns that take us from the
founding of stellar dynamics by Jeans and Eddington at the beginning
of the twentieth century to modern times. Chandrasekhar himself
contributed both fresh footpaths and blind alleys to this mazy route.

\section{Eddington and the Ellipsoidal Hypothesis}
\label{sec:eeh}

Eddington's (1915) paper that studies the ellipsoidal hypothesis is
one of his great ones. We can do no better than use Eddington's own
words:

\noindent {\it ``At any point of the system, the directions of the
  axes of the velocity ellipsoid determine three directions at right
  angles. The velocity ellipsoids thus define three orthogonal
  families of curves, each curve being traced by moving step by step
  always in the direction of an axis of the velocity ellipsoid at the
  point reached. These curves may be regarded as the intersections of
  a triply orthogonal family of surfaces, which we shall call the
  principal velocity surfaces. The axes of the velocity ellipsoid at
  any point are normals to the three principal velocity surfaces
  through any point''.}

In modern language, the theory of collisionless systems such as
galaxies begins with the Boltzmann equation:
\begin{equation}{\partial F \over \partial t} + {\bv} \cdot {\partial F
  \over \partial \bx} - {\partial \Phi \over \partial \bx} \cdot
{\partial F \over \partial \bv} =0,
\end{equation}
where $F$ is the phase space distribution function and $\Phi$ is the
gravitational potential.  At every point in the galaxy, we can define
a velocity dispersion tensor
\begin{equation}\sigma_{ij} = \langle (v_i - \langle v_i \rangle)(v_j
  - \langle v_j \rangle),
\end{equation}
where angled brackets denote averages over the distribution
function. The velocity dispersion tensor $\sigma_{ij}$ is real and
symmetric, and therefore by a well-known theorem in linear algebra has
mutually orthogonal eigenvectors. Eddington is asserting that the
integral curves of the eigenvectors provide a one-dimensional
foliation into surfaces, which he calls {\it the principal velocity
surfaces.}  We shall return to the assumptions underlying this
assertion shortly, as it is the precisely the point that troubled
Chandrasekhar.

Eddington then shows via Lagrange's equations that a steady state
distribution of stars moving in a gravitational potential $\Phi$
necessarily generates principal velocity surfaces that are confocal
quadrics. Labelling the quadric surfaces by $(\lambda,\mu, \nu)$,
these are recognised as ellipsoidal coordinates (e.g., Morse \&
Feshbach 1953). Eddington now proves two further theorems. First,
suppose that the distribution of velocities has exactly the
Schwarzschild (1908) or triaxial Gaussian form
\begin{equation} F \propto \exp \left( -{v_\lambda^2\over 2\sigma_\lambda^2} - {v_\mu^2\over
    2\sigma_\mu^2} - {v_\nu^2\over 2\sigma_\nu^2} \right),
\end{equation}
where ($v_\lambda,v_\mu,v_\nu$) are velocity components referred to
the locally orthogonal axes and ($\sigma_\lambda, \sigma_\mu,
\sigma_\nu$) are the semiaxes of the velocity ellipsoid. This is the
ellipsoidal hypothesis. Eddington showed that the only solutions for
the principal velocity surfaces are spheres. However, the
gravitational potential need not be spherical, but can take the
general form
\begin{equation}\Phi(r, \theta, \phi) = f(r) + {g(\theta) \over r^2}  + {h(\phi)
  \over r^2 \sin^2 \theta},
\end{equation}
where $f,g$ and $h$ are arbitrary functions of the indicated
arguments. These have sometimes been called Eddington potentials in
the astronomical literature.

Secondly, Eddington considered the more general case of a stellar
population with an arbitrary distribution of velocities. Under the
assumption of the existence of principal velocity surfaces, he showed
that the potential can take the general form in ellipsoidal
coordinates
\begin{equation}\Phi (\lambda, \mu, \nu) = {f(\lambda) \over (\lambda -\mu)(\lambda - \nu)} +
{g(\mu) \over (\mu - \lambda)(\mu - \nu)} + {h(\nu) \over (\nu
  -\mu)(\nu - \lambda)}.
\end{equation}
Eddington does not consider the fully triaxial case in detail, but he
does study the degenerations of the ellipsoidal coordinates into
spheroidal coordinates. Here, the stars have oblate or prolate density
distributions, the principal velocity surfaces are prolate or oblate
spheroids and the velocity dispersion tensor is in general
anisotropic. This was the first attempt to build galaxy models using
the separable potentials. Except in the spherical limit, Eddington
did not write down the form of the integrals of motion, leaving that
task to his student, G.L. Clark (1937).

Although Eddington's paper is not without its flaws, it turned out to
be remarkably prescient, anticipating developments over half a century
later.

\section{Chandrasekhar's Criticism}
\label{sec:cc}

In retrospect, Chandrasekhar's venture into stellar dynamics seems
both natural and brave. It is natural, as it is an obvious progression
of his interests in stellar structure and evolution. It is brave, as
it strays onto territory that Eddington had already made his own.  The
discipline had been founded by two people -- Eddington in his book
{\it Stellar Movements and the Structure of the Universe} published in
1914, and Jeans in his 1917 Adams Prize essay, published somewhat
later in 1919 as {\it Problems of Cosmogony and Stellar
  Dynamics}. Eddington and Jeans had dominated the subject over the
1920s, with fundamental contributions, including Jeans' theorem, the
equations of stellar hydrodynamics (sometimes called the Jeans'
equations), and Eddington's inversion formula for the distribution
function of a spherical galaxy.  Given Chandrasekhar's worsening
relationship with Eddington over these years, his incursions into this
field were almost inevitably opening up a second front.

Chandrasekhar (1939, 1940) announced his entry into the field with two
gigantic papers on the ellipsoidal hypothesis (summarised in Chapters
3 and 4 of {\it Principles of Stellar Dynamics} which themselves
occupy over a hundred pages). Right away, he detected an error in
Eddington's paper.  Chandrasekhar's criticism is worth quoting in
full:

\noindent
{\it ``The fallacy in Eddington's argumentation is clear. It is true
  that we can regard the directions of the principal axes of the
  velocity ellipsoid at any given point as being tangential to the
  three curves which intersect orthogonally at the point
  considered. But it is not generally true that we can regard these
  curves as the intersections of a triply orthogonal system of
  surfaces. Consequently, the notion of principal velocity surfaces
  introduces severe restrictions on the problem, which are wholly
  irrelevant and certainly unnecessary.''}

Here, Chandrasekhar is completely correct. Eddington assumed that the
eigenvectors of the velocity dispersion tensor are the tangent vectors
of a triply orthogonal system of surfaces. This is a sufficient, but
not a necessary, consequence of the orthogonality of the eigenvectors
of the dispersion tensor.  Eddington (1943) himself conceded as much
in his review of Chandrasekhar's book. Writing in the journal {\it
  Nature}, he stated:

\noindent
{\it ``Chandrasekhar rightly points out a fallacy in a theorem which I
  gave in 1915 and the correction makes the conclusion less general
  than has hitherto been assumed. But he does not take the
  opportunity of restating the position. Presumably it is still true
  that in a steady system with axial symmetry, the velocity surfaces
  are confocal quadrics and transverse star streaming is necessarily
  excluded, but there is no mention of this''}.

Where did Chandrasekhar's insight lead ? Chandrasekhar first somewhat
generalised the problem by asking for stellar dynamical models with
distribution functions $F$ of the form
\begin{equation}
F = F(Q),
\end{equation}
where $Q$ is a quadratic function of the velocities. The coefficients
are arbitrary functions of position. More formally,
\begin{equation}
Q = \bv \cdot {\bf M(\bx)} \cdot \bv + N(\bx),
\end{equation}
where ${\bf M}$ and $N$ are matrix and scalar functions of
position. This is a generalized ellipsoidal hypothesis, as $Q$ and
hence the phase space density $F$ is constant on ellipsoids in
velocity space.

Chandrasekhar proceeds by substituting his ansatz for the distribution
function into the Boltzmann equation and separating term by term in
the powers of velocity. He extracts a set of 20 partial differential
equations, which he reduces to 6 integrability conditions. Note that
Chandrasekhar does not impose the Poisson equation, as he is
interested in finding the conditions that a stellar population has a
distribution function of ellipsoidal form in an externally imposed
potential. He reaches a very surprising conclusion that {\it for
  stellar systems in a steady state, the potential $\Phi$ must
  necessarily be characterised by helical symmetry. The case of axial
  symmetry is included as a special case.}

In other words, using cylindrical polar coordinates ($R, \phi, z$),
Chandrasekhar asserts that the only solutions for the gravitational
potential compatible with the generalised ellipsoidal hypothesis are
\begin{equation}
\Phi = f(R, z + \alpha \phi),
\end{equation}
where $f$ is an arbitrary function of the indicated arguments and
$\alpha$ is a constant (the reciprocal of the pitch of the helix). The
integrals of motion are the energy $E$ and the generalisation of the
angular momentum component, namely
\begin{equation}
I = p_\phi - \alpha p_z
\end{equation}
where $p_\phi$ and $p_z$ are the canonical momenta conjugate to $\phi$
and $z$.  Chandrasekhar then notes that such a potential can have
bound orbits only if it is axisymmetric ($\alpha =0$) and so he
reaches his final conclusion.  {\it For stellar systems with
  differential motions, which are in steady states and of finite
  spatial extent, the potential $\Phi$ must necessarily be
  characterized by axial symmetry}.

This is a strong claim (we shall shortly see that, like Eddington's
work, it is not entirely correct).  A surprising aspect is that,
having realised that Eddington had introduced unnecessarily
restrictive assumptions into the problem, Chandrasekhar is not
troubled by that fact that his more general approach finds fewer
solutions than Eddington -- and indeed doesn't find the solutions with
quadric principal velocity surfaces at all!  Even more curiously,
Chandrasekhar recognises that the phase space distribution $F$ is an
integral of motion, quoting Whittaker's (1936) book on {\it Analytical
  Dynamics} as a reference. He therefore knows that his problem is
exactly equivalent to seeking all potentials that admit integrals of
motion quadratic in the velocities.  But, this problem is also
(partly) solved in Whittaker's book, which provides a derivation of
the separable potentials in spheroidal coordinates, though not
ellipsoidal, from the assumption of quadratic integrals.

A new result of Chandrasekhar is that he provides an explicit
counter-example to Eddington's assumption. The helically symmetric
systems indeed remain the only ones known to us which do not possess
mutually orthogonal principal velocity surfaces, but do satisfy the
ellipsoidal hypothesis. They are not of much astrophysical interest as
they do not resemble galaxies, but they remain of considerable
intellectual interest.

Another insight of Chandrasekhar that has proved its worth is his
repeated emphasis on the principal of equivalence. By this, he means
that if several different models can be found sharing the same
gravitational potential, then a more complex model that does not
satisfy the ellipsoidal hypothesis can be built by weighted linear
superposition. This idea has often been exploited in modern times to
build realistic models by superpositions of analytic distribution
functions (e.g., Fricke 1952, Dejonghe 1989, Emsellem et al. 1999).

\section{A Modern Approach}
\label{sec:mod}

Let us now state and give the solution to Chandrasekhar's problem anew
from the point of view of a modern dynamicist. Jeans' theorem tells us
that the distribution function of a collisionless system depends only
on the globally defined, isolating, integrals of motion. It therefore
follows that $Q$ must be an integral of motion.  Chandrasekhar's
problem is exactly equivalent to identifying all those potentials that
admit integrals of motion at most quadratic in the velocities. This is
a problem of widespread interest in both mathematics and physics, with
an enormous literature and history.

Integrals of motion that are linear in the velocities always result
from geometric symmetries of space. This is sometimes called Noether's
theorem (see e.g., Landau \& Lifshitz 1976, Arnold 1978). It follows
from the fact that the Lagrangian is invariant with respect to the
corresponding transformations, which are linear in the generators of
the Euclidean group of symmetries. Examples include the invariance of
the angular momentum component $p_\phi$ in axisymmetric potentials
$\Phi(R,z)$, and the invariance of the linear momentum component $p_z$
in translationally invariant potentials $\Phi(x,y)$. Chandrasekhar's
helical solution is the most general possible, with rotationally and
translationally invariant potentials given by the limits $\alpha
\rightarrow 0$ and $\alpha \rightarrow \infty$ respectively.

Integrals of motion that are quadratic in the velocities always result
from separability of the Hamilton-Jacobi equation in some coordinate
system.  Many authors discovered some or all of the potentials for
which the Hamilton-Jacobi equation is separable in the confocal
ellipsoidal coordinates or their degenerations (e.g., Eddington 1915,
Weinacht 1924, Whittaker 1936, Clark 1937, Eisenhart 1948, Lynden-Bell
1962). These systems possess integrals of the motion quadratic in the
velocities by construction, as the Hamilton-Jacobi equation only has
such terms in it! The fact that separability of the Hamilton-Jacobi
equation is both a necessary and sufficient condition is a much more
difficult result to prove. It was done for the first time in Makarov
et al. (1967).

Although written from the viewpoint of particle physicists, Makarov et
al. (1967) follow essentially the same route as Chandrasekhar in
Chapter 3 of {\it Principles of Stellar Dynamics}. That is, they ask
for the Poisson bracket of the integral of motion $Q$ with the
Hamiltonian $H$ to vanish. This is mathematically identical to
requiring that $Q$ satisfy the collisionless Boltzmann equation, as
Chandrasekhar did. The main difference is that Makarov et
al. substantially simplify $Q$ by rotations and translations, before
requiring that $Q$ commute with the Hamiltonian $H$. This
considerably reduces the mathematical complexity of the problem,
enabling them to find all possible solutions (including the separable
ones that Chandrasekhar had missed).

Before passing to later developments, it is worth remembering that
Chandrasekhar and Eddington had disagreed over the white dwarf stars
and the endpoints of stellar evolution (see Vibert Douglas 1956, Wali
1996, Chandrasekhar 1988 for various perspectives on this affair). In
retrospect, it is clear that Eddington behaved badly over the white
dwarfs, not so much because he was wrong -- that can (and should)
happen to every scientist! -- but because he used his seniority to
stifle the work of a younger colleague.

Is it possible that Chandrasekhar, hurt by the reception of what would
ultimately prove to be a Nobel Prize winning achievement, was unable
to appreciate fully the advantages in Eddington's approach in stellar
dynamics?  True, he had detected an error in Eddington's (1915) paper,
but Eddington in the end saw closer to the truth of the matter in
stellar dynamics. Eddington introduced a hypothesis -- the principal
velocity surfaces -- that was not strictly-speaking necessary and
would ultimately be discarded by later scientists. But, it proved to
be a physically fruitful hypothesis that led Eddington to an important
class of models. Consequently, later developments were to build more
on Eddington's work than Chandrasekhar's, as we will now see.

\section{Later Developments}
\label{sec:lated}

Further advances had to wait till the late fifties and early sixties,
when the subject was revived by Lynden-Bell (1962) with a particularly
original investigation. Rather than starting from an assumption that
the integrals are quadratic in the velocities, Lynden-Bell permitted
the integrals to have any form (polynomial or transcendent). Instead,
he assumed that the steady-state is one of a set through which the
system may secularly evolve whilst preserving the existence of the
integrals of motion. This led to the enumeration of all potentials
with such isolating integrals -- prominent among them being the
separable potentials in ellipsoidal coordinates and their
degenerations. At the time, the flattening of elliptical galaxies was
believed to be caused primarily by rotation rather than velocity
anisotropy. Hence, the application of the potentials to galaxies
remained unexplored in the West.

This was not true in the former Soviet Union, as a remarkable and
sadly neglected paper by Kuzmin (1957) -- citing the influences of
Eddington (1915) and Clark (1937) -- had already used the separable
potentials in spheroidal coordinates to build an oblate, axisymmetric
model of the Galaxy. Kuzmin (1973) was also the first to write down
the fully triaxial case, and study its orbital structure, identifying
the 4 characteristic classes of orbits: box, inner and outer long axis
tubes and short axis tubes (see e.g., Binney \& Tremaine 1987). These
models became well-known in the West only after they had been
re-discovered and extended by de Zeeuw (1985). Kuzmin (1973) and de
Zeeuw (1985) showed that an ellipsoidally stratified model with
density
\begin{equation} \rho = {\rho_0 \over (1 + m^2)^2}, \qquad\qquad m^2 = {x^2\over
  a^2} + {y^2\over b^2} + {z^2\over c^2}
\end{equation}
possesses an exactly separable gravitational potential in confocal
ellipsoidal coordinates. The easiest way to demonstrate this is by
making use of the methods and formulae in Chandrasekhar's (1968)
finest and most beautiful book, {\it Ellipsoidal Figures of
  Equilibrium}. This is an important result as it showed that
realistic and phyically motivated models of elliptical galaxies could
be built from separable potentials.  De Zeeuw also demonstrated a
number of beautiful properties of these models, including the
classification of their orbits in integral and action
space\footnote{By now, these potentials had come to be known as
  St\"ackel potentials in the astronomical literature. This seems
  unwarranted.  First, it is poor practice in physics to associate a
  name with an equation if a perfectly adequate descriptive term
  exists. On these grounds alone, the term 'separable potential' is
  preferable to 'St\"ackel potential'. And, second, there is no reason
  to associate the name of Paul St\"ackel with coordinate systems and
  potentials that he never wrote down!  St\"ackel was a prominent
  differential geometer, latterly Professor of Mathematics at
  Heidelberg. In his {\it Habilitationschrift} in 1891 at Halle,
  St\"ackel wrote down the condition for the Hamilton-Jacobi equation
  to separate in a given coordinate system on a general Riemannian
  manifold in the form of the vanishing of a determinant (which has
  reasonably enough come to be called the St\"ackel
  determinant). St\"ackel did not derive the coordinate systems in
  Euclidean 3-space for which his determinant vanishes, far less the
  form of the separable potentials in these coordinates. This work was
  left to Weinacht (1924) and Eisenhart (1948). In fact, St\"ackel's
  result is limited, as it does not even provide a comprehensive test
  for separability. St\"ackel's determinant for a separable system
  only vanishes if it is written down in the separable coordinate
  system itself. The finding of a general criterion for identifying
  whether a potential is separable in some coordinate system remains
  an outstanding research problem.}.  This led to a flowering of
interest in the models, as evidenced by the papers in {\it IAU
  Symposium 127: The Structure and Dynamics of Elliptical Galaxies}
(de Zeeuw 1987). This can be seen as the culmination of over seventy
years of astronomical research on the subject, from Eddington, through
Chandrasekhar, Lynden-Bell and Kuzmin to modern times.

Even though their mass density falls off faster than the luminosity
density of giant ellipticals, and even though they are cored in the
central parts rather than cusped, the separable models still occupy a
special place in modern galactic dynamics.  This is because the
orbital structure of the models is generic for all flattened triaxial
systems without figure rotation. Although the models do not contain
any irregular or chaotic orbits, for many applications in galactic
dynamics, this is unimportant, as the fraction of phase space occupied
by truly irregular orbits is believed to be small (Goodman \&
Schwarzschild 1981).

\begin{figure}
\begin{center}
\epsfig{file = 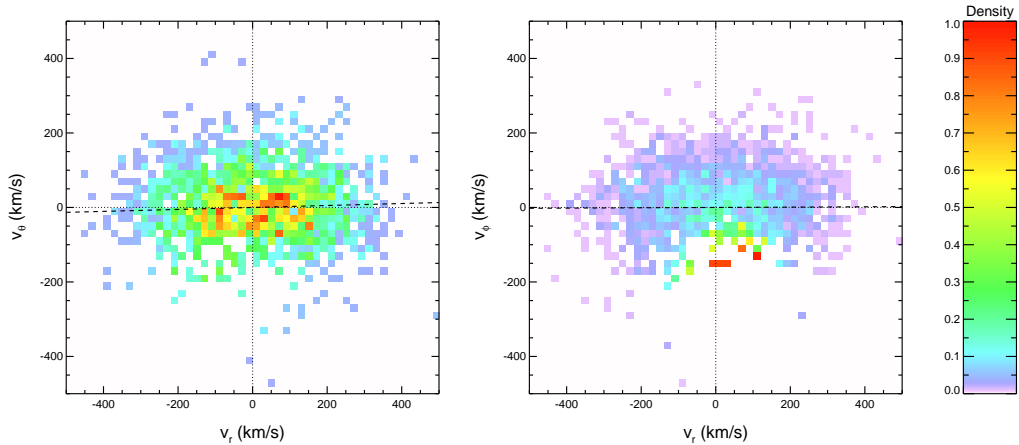,height=6cm}
\caption{The efficiency corrected velocity distributions in the
  $(v_r,v_\theta)$ and $(v_r,v_\phi)$ planes for the sample of 1,600
  subdwarfs with $1~\mbox{kpc} < |z| < 4~\mbox{kpc}$.  The dashed
  lines show the orientation of the tilts, which are very close to
  spherical alignment. The apparent non-Gaussianity in the
  $(v_r,v_\phi)$ distribution is due to the variation of the
  efficiency correction across this plane. [From Smith et
  al. (2009b)].}
\end{center}
\end{figure}

\section{The Alignment of the Velocity Dispersion Tensor}
\label{sec:tilt}

Modern interest in the subject (e.g., Smith, Evans \& An 2009a,b,
Binney \& McMillan 2011) has been given additional impetus by
large-scale photometric and spectroscopic surveys of hundreds of
thousands of stars in the Milky Way Galaxy itself.  If proper motions
are also available, then this raises the possibility that all the
components of the velocity dispersion tensor can be computed directly
from the data. There have been a number of interesting recent attempts
to do this, both for halo and disk populations. Although sample sizes
are presently still small, and distance errors a serious hazard,
matters will substantially improve in the next few years.

For example, the {\it Sloan Digital Sky Survey} (SDSS, York et
al. 2000) carried out repeated photometric measurements in an
equatorial stripe, known as Stripe 82, primarily with the aim of
supernova detection.  Bramich et al. (2008) then provided a public
archive of light-motion curves in Stripe 82 complete down to magnitude
$21.5$ in the $u$, $g$, $r$ and $i$ photometric bands, and to
magnitude $20.5$ in $z$. This reaches almost 2 magnitudes fainter than
the SDSS/USNO-B catalogue (Munn et al. 2004), making it the deepest
large-area photometric and astrometric catalogue available. Smith et
al. (2009a,b) extracted a sample of $\sim$1,600 halo subdwarf stars
via a reduced proper motion diagram. Their radial velocities are
calculated from the SDSS spectra and their distances are estimated
from photometric parallaxes, thus giving the full phase space
information. Although the sample is not kinematically unbiased, the
detection efficiency can be calculated and corrections made for any
biases.

Figure~1 shows the velocity distributions of the SDSS Stripe 82
subdwarfs. These halo stars lie at Galactocentric cylindrical polar
radii between 7 and 10 kpc, and at depths of 4.5 kpc or less below the
Galactic plane. The good alignment of the velocity ellipsoid in
spherical polars is already apparent from the velocity distributions
in the ($v_r, v_\theta$ ) and ($v_r, v_\phi$) planes. Smith et
al. find that the velocity dispersion tensor of the halo subdwarfs has
semiaxes ($\sigma_r, \sigma_\phi, \sigma_\theta$) = ($143 \pm 2, 82
\pm 2, 77 \pm 2$) kms${}^{-1}$.  The misalignment from the spherical
polar coordinate surfaces can then be described by the correlation
coefficients and the tilt angles using
\begin{equation}
{\rm Corr}[v_i,v_j] = 
\frac{ \sigma_{ij}^2 }{ (\sigma_{ii}^2\sigma_{jj}^2)^{1/2} },
\end{equation}
and
\begin{equation}
\tan(2\tilt) = 
\frac{ 2\sigma_{ij}^2 }{ \sigma_{ii}^2 - \sigma_{jj}^2 }.
\label{eq:tiltdef}
\end{equation}
The tilt of the velocity ellipsoid with respect to the spherical polar
coordinate system is found to be consistent with zero for two of the
three tilt angles, and very small for the third. Specifically, Smith
et al find:
\begin{eqnarray}
{\rm Corr} [v_r, v_\theta] &= 0.078\!\pm\!0.029,\quad&
\tiltrt= 3\fdg4\!\pm\!1\fdg3,
\nonumber \\
{\rm Corr} [v_r, v_\phi] &= -0.028\!\pm\! 0.039,\quad&
\tiltrp = -2\fdg2\!\pm\!3\fdg3,
\\
{\rm Corr} [v_\phi, v_\theta] &= -0.087\!\pm\! 0.047,\quad&
\tiltpt = -37\fdg4\!\pm\!20\fdg4.
\nonumber
\end{eqnarray}
In Eddington's language, these stars have spherical principal velocity
surfaces to an excellent approximation. In a slight extension of the
earlier results of Eddington (1915) and Chandrasekhar (1939), Smith et
al. (2009b) prove that: {\it If the potential is nonsingular, it is a
  sufficient condition for a spherical symmetric potential that one of
  the non-degenerate eigenvectors of the velocity dispersion tensor is
  aligned radially everywhere.}

Of course, Smith et al. (2009b) did not demonstrate that the velocity
dispersion tensor is aligned everywhere in spherical polar
coordinates. They showed that the alignment is very close to spherical
for halo subdwarfs at heliocentric distances of $< 5~\mbox{kpc}$ along
the $\sim 250~\mbox{deg$^2$}$ covered by SDSS Stripe 82. Nonetheless,
they argued that this is still a striking and unexpected result over a
range of Galactic locations that provides a new line of attack on the
awkward question of the shape of the Milky Way's dark halo.  Binney \&
McMillan (2011) concur that local measurements are not enough to
constrain the shape of the Galaxy's potential. Further work on the
alignment of the velocity ellipsoid of halo populations is highly
desirable.

\begin{figure}
\begin{center}
\epsfig{file=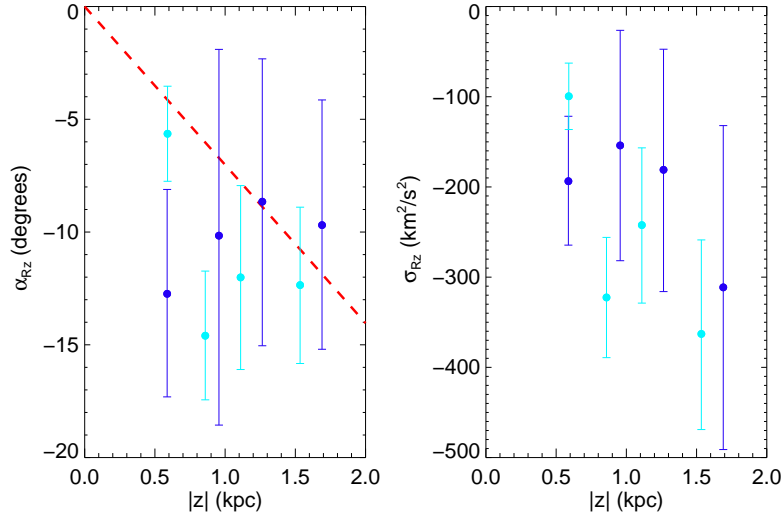,height=8cm}
\caption{The variation of $\sigma_{Rz}$ and the corresponding angle
  $\alpha_{Rz}$ as a function of height from the plane.  The dashed
  red line is the assumed halo tilt (i.e. aligned in spherical
  polars). The blue and cyan points correspond to disc stars with
  metallicities $-0.8 \le \feh \le -0.5$ and $\feh \ge -0.5$,
  respectively.  [From Smith et al. (2011)].}
\end{center}
\end{figure}

By contrast, the behaviour of the velocity ellipsoid of disk
populations has been more widely studied, not least because of its
importance for calculations of the asymmetric drift and the Oort
Limit.  Based on evidence from orbit integrations, Binney \& Tremaine
(1987) suggest that the tilt may lie midway between spherical and
cylindrical polar alignment. This is also the expectation from models
based on potentials separable in spheroidal coordinates (Statler
1989).  There have been three recent determinations directly from data
by Siebert et al. (2008), Fuchs et al. (2009) and Smith, Evans \&
Whiteoak (2011).

Siebert et al. (2008) extracted 763 red clump stars from the {\it
  Radial Velocity Experiment} dataset (RAVE, Zwitter et al. 2008),
spanning a distance interval from the Sun of 500 to 1500 pc. The tilt
of the velocity ellipsoid of stars so close to the Galactic plane is
affected both by the structure of the Galactic disk and and the
flattening of the dark halo.  Siebert et al. find that the velocity
ellipsoid is tilted towards the Galactic plane with an inclination of
$7\fdg3 \pm 1\fdg8$. This is entirely consistent with alignment in
spherical polar coordinates. Siebert et al. compare this value to
computed inclinations for two mass models of the Milky Way. The
measurement is consistent with a short scalelength of the stellar disc
($\approx 2$ kpc) if the dark halo is oblate or with a long
scalelength ($\approx 3$ kpc) if the dark halo is spherical or
prolate.

Fuchs et al. (2009) used an enormous sample of $ \sim 2$ million M
dwarfs derived from the {\it Sloan Digital Sky Survey} Data Release 7
(Abazajian et al. 2008). Although the proper motions and photometric
distances of these stars are available, unfortunately the radial
velocities are not. Fuchs et al. estimated the radial velocities via
the method of deprojection of proper motions. They found an
anomalously large tilt reaching an inclination of $20^\circ$ at
heights above the Galactic plane of 800 pc, whereas spherical
alignment would predict an inclination of $\approx 5^\circ$.  McMillan
\& Binney (2009) have argued that this surprisingly large value may be
spurious, a consequence of correlations between velocities and
positions of stars, which renders the method of deprojection invalid.

Finally, Smith et al. (2011) again use the very deep light-motion
catalogue for Stripe 82 (Bramich et al. 2008) to extract a sample of
disk stars, complete with radial velocities from SDSS spectra and
photometric metallicities. These stars are confined to a narrow range
of cylindrical polar radius between $7 \le R \le 9$ kpc. However,
there are enough stars to split the data into three ranges in
metallicity ($-1.5 \le \feh \le -0.8, -0.8 \le \feh \le -0.5$ and
$-0.5 \le \feh$), and for each metallicity bin to divide the data into
four ranges in z ($0 \le |\rm{z}| \le 0.8$, $0.8 \le |\rm{z}| \le
1.1$, $1.1 \le |\rm{z}| \le 1.5$ and $1.5 \le |\rm{z}| \le 2.2$
kpc). This gives around 500 to 800 stars per bin. The variation with
height and metallicity is shown in Figure 2.  The dotted line
corresponds to what we would expect for a velocity ellipsoid aligned
in spherical polar coordinates. The metal-rich and medium-metallicity
stars are arguably consistent with the dotted line, and hence
consistent with the result of Siebert (2008). In general, the stars in
the lowest metallicity bins (not plotted in Figure 2) exhibit tilt
angles which are larger than this, albeit with very noisy error bars.

Fortunately, the very-near future sees the dawning of the Age of
Precision Astrometry.  The GAIA satellite (e.g., Gilmore 2007) will
provide tangential velocities for 44 million stars and distances for
21 million stars with an accuracy better than 1 per cent. There is
therefore a realistic prospect that the behaviour of the velocity
ellipsoid for both disk and halo populations over a swathe of
locations in the Milky Way Galaxy will be known shortly.

\section*{Acknowledgements}
I am grateful to Donald Lynden-Bell and Tim de Zeeuw for first
introducing me to this fascinating subject. I thank too my
collaborators Jin An, Martin Smith and Hannah Whiteoak for their help
and insight, as well as permission to report results in advance of
publication.

\label{lastpage}
\end{document}